\documentclass[journal=jacsat,manuscript=article]{achemso}

\usepackage[version=3]{mhchem} 
\usepackage{amsmath}
\usepackage{amssymb}
    \usepackage{amsthm}
    \usepackage{amsfonts}
    \usepackage{braket}
\usepackage{gensymb}
\usepackage{bm}
\usepackage{subcaption}
\usepackage{graphicx}
\usepackage{color}
\DeclareCaptionFormat{custom}
{%
    \textbf{#1#2}#3
}
 \usepackage{hyperref}
 \hypersetup{
     colorlinks=true,
     linkcolor=blue,
     filecolor=blue,
     citecolor = black,      
     urlcolor=black,
     }

\author{Hao Wu}
\affiliation[BIT]
{Center for Quantum Technology Research and Key Laboratory of Advanced Optoelectronic Quantum Architecture and Measurements (MOE), School of Physics, Beijing Institute of Technology, Beijing 100081, China}
\alsoaffiliation[baqis]
{Beijing Academy of Quantum Information Sciences, Beijing 100193, China}
\altaffiliation{These authors contributed equally: Hao Wu, Tong Li}

\author{Tong Li}
\affiliation[BIT]
{Center for Quantum Technology Research and Key Laboratory of Advanced Optoelectronic Quantum Architecture and Measurements (MOE), School of Physics, Beijing Institute of Technology, Beijing 100081, China}
\alsoaffiliation[baqis]
{Beijing Academy of Quantum Information Sciences, Beijing 100193, China}
\altaffiliation{These authors contributed equally: Hao Wu, Tong Li}

\author{Zhang-Qi Yin}
\affiliation[BIT]
{Center for Quantum Technology Research and Key Laboratory of Advanced Optoelectronic Quantum Architecture and Measurements (MOE), School of Physics, Beijing Institute of Technology, Beijing 100081, China}

\author{Jiyang Ma}
\email{mjy@bit.edu.cn}
\affiliation[BIT]
{Center for Quantum Technology Research and Key Laboratory of Advanced Optoelectronic Quantum Architecture and Measurements (MOE), School of Physics, Beijing Institute of Technology, Beijing 100081, China}
\alsoaffiliation[baqis]
{Beijing Academy of Quantum Information Sciences, Beijing 100193, China}

\author{Xu-Ri Yao}
\affiliation[BIT]
{Center for Quantum Technology Research and Key Laboratory of Advanced Optoelectronic Quantum Architecture and Measurements (MOE), School of Physics, Beijing Institute of Technology, Beijing 100081, China}
\alsoaffiliation[baqis]
{Beijing Academy of Quantum Information Sciences, Beijing 100193, China}

\author{Bo Zhang}
\affiliation[BIT]
{Center for Quantum Technology Research and Key Laboratory of Advanced Optoelectronic Quantum Architecture and Measurements (MOE), School of Physics, Beijing Institute of Technology, Beijing 100081, China}
\alsoaffiliation[baqis]
{Beijing Academy of Quantum Information Sciences, Beijing 100193, China}

\author{Mark Oxborrow}
\affiliation[Imperial]
{Department of Materials, Imperial College London, South Kensington, SW7 2AZ London, United Kingdom}

\author{Qing Zhao}
\affiliation[BIT]
{Center for Quantum Technology Research and Key Laboratory of Advanced Optoelectronic Quantum Architecture and Measurements (MOE), School of Physics, Beijing Institute of Technology, Beijing 100081, China}
\alsoaffiliation[baqis]
{Beijing Academy of Quantum Information Sciences, Beijing 100193, China}

\title{Towards simultaneous coherent radiation in the visible and microwave bands with doped molecular crystals}

\begin{document}







\begin{abstract}
  Coherent sources exploiting the stimulated emission of non-equilibrium quantum systems, i.e. gain media, have proven indispensable for advancing fundamental research and engineering. The operating electromagnetic bands of such coherent sources have been continuously enriched for increasing demands. Nevertheless, for a single bench-top coherent source, simultaneous generation of radiation in multiple bands, especially when the bands are widely separated, present formidable challenges with a single gain medium. Here, we propose a mechanism of simultaneously realizing the stimulated emission of radiation in the visible and microwave bands, i.e. lasing and masing actions, at ambient conditions by utilizing photoexcited singlet and triplet states of the pentacene molecules that are doped in \textit{p}-terphenyl. The possibility is validated by the observed amplified spontaneous emission (ASE) at 645 nm with a narrow linewidth around 1 nm from the pentacene-doped \textit{p}-terphenyl crystal used for masing at 1.45 GHz and consolidated by a 20-fold-lower threshold of ASE compared to the reported masing threshold. The overall threshold of the pentacene-based multiband coherent source can be optimized by appropriate alignment of the pump-light polarization with the pentacene's transition dipole moment. Our work not only shows a great promise on immediate realization of multiband coherent sources but also establishes an intriguing solid-state platform for fundamental research of quantum optics in multiple frequency domains.
\end{abstract}

\section{Introduction}

Coherent sources that capable of generating coherent electromagnetic radiation have served as crucial ingredients enabling numerous breakthroughs in the fields of physical, environmental, and biological sciences. A well-established approach to achieve coherent electromagnetic radiation is to exploit the stimulated emission process\cite{einstein1916quantentheorie} induced by interactions between electromagnetic fields and matters. Stemming from the discovery of the stimulated emission, optical lasers spanning from the ultraviolet to the near-infrared region, together with their forerunners and microwave analogs, masers, have become indispensable coherent sources for applications in telecommunication\cite{nagourney2014quantum}, metrology\cite{hall2001ultrasensitive,arias2005metrology}, sensing\cite{hahl2022magnetic,wu2022enhanced,jiang2021floquet}, machining\cite{dubey2008laser} and quantum information\cite{lamas2001stimulated,breeze2017room}. In addition, the recent development of terahertz\cite{chevalier2019widely} and X-ray\cite{bostedt2016linac} lasers has offered alluring promise of shedding light on astronomical observations, spectroscopy and structural biology. The rich variety of applications require distinct coherent sources operational across the extremely wide electromagnetic spectrum from microwave to X-rays. Even though the underlying mechanism of those coherent sources is the same (i.e. the stimulated emission), their practical realizations are completely different in terms of the components, structures and scale, rendering simultaneous generation of coherent radiation in widely separated spectrum bands with a single source hitherto unattainable on the bench top. 

Gain media, constituted by the quantum systems with non-degenerate energy levels, are the core determining radiation wavelengths of the stimulated emission, therefore, vital for realizing multiband coherent sources. It is not scarce for quantum systems to possess multiple transitions across different spectrum bands by their intrinsic properties (e.g. large quantum numbers\cite{Gallagher2006}) and/or external manipulations with electric/magnetic fields\cite{condon1935gh}. Ruby (chromium ions-doped aluminum oxide) is a representative gain medium which can be employed for both solid-state masers\cite{strayer1987performance} and lasers\cite{maiman1960stimulated}, i.e. capable of generating coherent radiation in the microwave and visible regions, while the vastly different experimental setups, especially the cryogenic and magnetic-field requirements for ruby masers, have resulted in no demonstration of simultaneous maser and laser actions of ruby to date. More recently, the negatively charged nitrogen-vacancy defects (NV$^{-}$) in diamond have been demonstrated to be promising solid-state maser\cite{breeze2018continuous} and laser\cite{savvin2021nv} gain media at ambient conditions. However, the preparations and material properties of the NV$^{-}$ diamonds employed for the maser and laser applications are rather different. The diamonds were synthesized via the routes of high pressure high temperature (HPHT) and chemical vapor deposition (CVD) for the NV$^{-}$ laser and maser, respectively, which gives rise to the different concentrations of the doped nitrogen atoms and NV$^{-}$. The concentration of the gain media, i.e. NV$^{-}$, required for the laser action is 1.4 p.p.m. which is about four folds higher than that (0.36 p.p.m.) used for the NV$^{-}$ maser. With respect to the performance of the NV$^{-}$ diamond based coherent sources, the relatively weak output ($\sim$1 pW) of the NV$^{-}$ maser\cite{breeze2018continuous} and the broad spectral linewidth (20 nm) of the NV$^{-}$ laser\cite{savvin2021nv} still need to be substantially improved for practical considerations. In addition to the solid-state systems, due to the rich energy structures, the stimulated emission of radiation ranging from the microwave to infrared (IR) domain have been observed in vapor of Rydberg atoms\cite{moi1983rydberg,grischkowsky1977atomic} but their ability of simultaneous generation of coherent radiation in multiple wavelength domains is still not evident. Moreover, the limited number of atoms has also restricted the power of such coherent sources. The output power of the Rydberg masers has been reported to be up to 10 pW\cite{moi1980heterodyne} and the pulse energy of a Rydberg IR laser was approximately 1 $\mu$J\cite{grischkowsky1977atomic}.

Compared to the gain media mentioned above, doped molecular crystals combines the key advantages of the inorganic solid-state systems (e.g. robustness and ease of integration) with rich opportunities of tuning the energy levels of the quantum systems, like Rydberg atoms, through bottom-up engineering of the guests and host matrices\cite{bayliss2020optically,bayliss2022enhancing}. While sustaining the satisfying optical, magnetic and electronic properties, the doping concentrations of molecular crystals are tunable up to tens of thousands p.p.m.\cite{lubert2018identifying} implying the great potential of realizing powerful coherent sources with such gain media. Among numerous doped molecular crystals, pentacene-doped \textit{p}-terphenyl (Pc:Ptp) has attracted extensive attention especially in the last decade. Pc:Ptp is a low-cost and easy-fabricated single organic crystal, which can be produced in bulk with a high quality\cite{cui2020growth}. Due to the doping structure, the functional dopants, i.e. pentacene molecules, are well protected by the matrix of \textit{p}-terphenyl, which not only possess great steadiness and durability but also emerge appealing photophysical properties (which are lacking in neat pentacene) well suited for multidisciplinary applications, such as photovoltaics\cite{lubert2018identifying}, dynamic nuclear polarization (DNP)\cite{takeda2003studies} and quantum information processing\cite{breeze2017room,wu2021bench,kothe2021initializing}. In particular, Pc:Ptp is the first, and so far the only doped molecular crystal capable of masing at room temperature in Earth’s field\cite{oxborrow2012room} and the superior maser output at a level of milliwatt (i.e. 10$^{9}$ times higher than that of NV$^{-}$ masers) has yet to be surpassed by other room-temperature maser gain media. The maser application as well as the recent applications mentioned above mainly exploits the photoexcited triplet states of pentacene in \textit{p}-terphenyl. It is worth noting that the singlet states of pentacene are also intriguing due to their photoluminescent properties, owing to which the stimulated emission of radiation in the visible region manifesting as amplified spontaneous emission (ASE) have been observed\cite{wang2009doped,zhao2019high} implying the potential of the pentacene molecules to be suitable gain media for lasing as well. However, since the ASE phenomena were observed with pentacene doped in trans-1,4-distyrylbenzene (trans-DSB)\cite{wang2009doped} and 1,4-bis(2-cyano styryl)benzene (2-CSB)\cite{zhao2019high}, respectively, instead of \textit{p}-terphenyl, and the dynamics of pentacene’s triplet spins can be modulated by the host effects\cite{ong1994peculiar}, the suitability of these two host matrices for the maser action of pentacene remains elusive.

In this work, we investigate the optical emission properties of the Pc:Ptp crystals employed in the previous maser studies\cite{wu2022enhanced,wu2021bench,wu2020room} and first demonstrate the ASE process in Pc:Ptp at 645 nm, where molecular crystals rarely reached, under the experimental conditions identical to those required for the maser action. Combining spectral analysis with nanosecond time-resolved characterizations, the ASE properties of Pc:Ptp are systematically studied at room temperature. The obtained ASE spectrum shows an extremely narrow linewidth around 1 nm that is the narrowest among the reported molecular crystals revealing ASE behaviors. Since ASE is a prerequisite for lasing, our results provide solid evidence that Pc:Ptp can serve as a solid-state multiband gain medium for emitting coherent microwave and visible light simultaneously at ambient conditions.

\section{Results}

\subsection{Structural and optical properties of Pc:Ptp}

\begin{figure}[htbp!]
    \includegraphics[scale = 0.75]{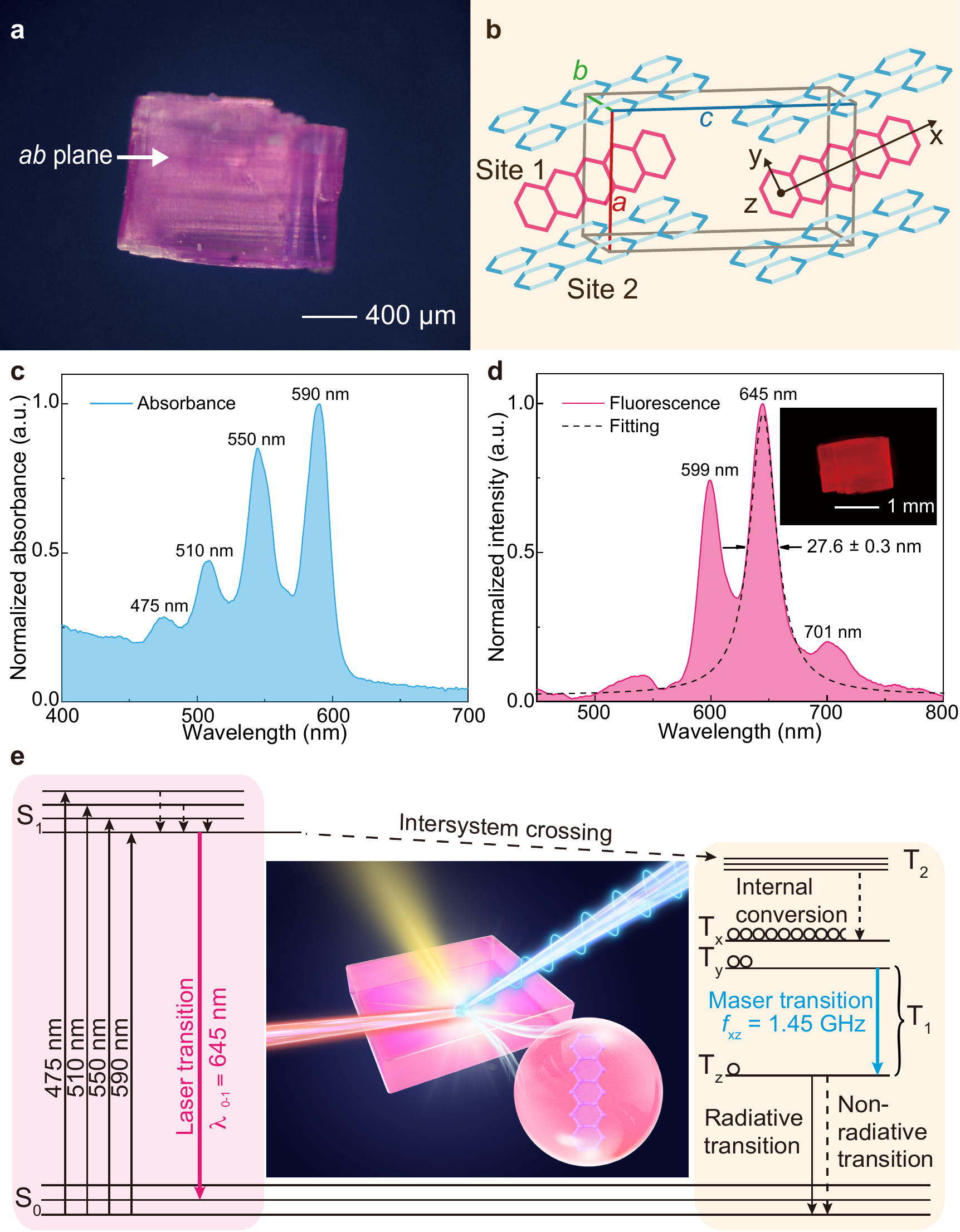}
    \caption{\textbf{Material characterizations of Pc:Ptp and mechanism of simultaneous coherent radiation.} \textbf{a} Optical microscopic image of a Pc:Ptp crystal under white-light illumination. The cleavage plane, i.e. \textit{ab} plane, of the crystal is labelled. \textbf{b} Crystal structure of the host matrix, \textit{p}-terphenyl (blue) with substitutionally doped pentacene molecules (pink). The two inequivalent doping sites as well as the molecular axes of pentacene are labelled. \textbf{c} UV/vis absorption and \textbf{d} Fluorescence spectra of Pc:Ptp with all characteristic peaks labelled. The linewidth, i.e. FWHM of the strongest fluorescent peak is determined by a Lorentzian fitting. Inset: optical microscopic image of a fluorescent Pc:Ptp with the excitation light filtered. \textbf{e} Proposed mechanism of simultaneous lasing and masing actions in Pc:Ptp exploiting the transitions of stimulated emission highlighted in the pentacene's singlet (pink) and triplet (orange) manifolds.}
    \label{fig:1}
\end{figure}

Throughout the study, pink Pc:Ptp single crystals, shown in Fig.~\ref{fig:1}a, with a doping concentration of 1000 p.p.m. was used, which is the same as that employed for room-temperature masers\cite{wu2022enhanced,wu2021bench,wu2020room}. The relatively high doping concentration allowing sufficient pentacene molecules for microwave and optical gains arises from the similar molecular packing coefficients of pentacene and \textit{p}-terphenyl which favors the formation of solid solutions with these two organic substances\cite{yu1982electron,kitaigorodsky2012molecular}. The molecular packing coefficient $k$ can be expressed as $k=(z\times V_{0})⁄V$, where $z$ is the number of molecules in the unit cell, $V_{0}$ and $V$ are the volumes of the molecule and unit cell, respectively. The $k$ values of pentacene and \textit{p}-terphenyl have been determined to be 0.743 and 0.751 that brings them adjacent to the closest packing condition where $k$= 0.74\cite{yu1982electron}. As shown in Fig.~\ref{fig:1}b, at room temperature, pentacene molecules are substitutionally doped in the monoclinic unit cell of \textit{p}-terphenyl with two inequivalent doping sites \cite{lang2007dynamics}. The structural properties of Pc:Ptp crystals are dominated by the host matrix, i.e. \textit{p}-terphenyl. Therefore, Pc:Ptp also possesses an intrinsic laminar structure with a cleavage (001) (i.e. \textit{ab}) plane\cite{cui2020growth} where the molecules stand (see Fig.~\ref{fig:1}b). The ‘head-to-head’ packing against the \textit{ab} plane results in the weaker intermolecular interactions compared to the $\pi-\pi$ interactions along the molecular xy plane where x and y denote the long and short axes of the molecules, respectively. The cleavage property facilitates the fabrication of Pc:Ptp crystals with distinguishable crystal planes. As exhibited in Fig.~\ref{fig:1}a, the large crystal facet can be straightforwardly determined to be the (001) plane due to the obvious delamination shown on the edge.

The optical properties of Pc:Ptp were investigated by measuring its absorption and fluorescence spectra as illustrated in Fig.~\ref{fig:1}c and \ref{fig:1}d. It can be found that the doped crystal reveals explicit absorption at wavelengths of 475, 510, 550 and 590 nm, which correspond to the characteristic transitions between pentacene’s excited and ground singlet states\cite{cui2020growth,bogatko2016molecular} as depicted in Fig.~\ref{fig:1}e. According to the absorption spectrum, the highest absorbance locates at 590 nm was referred to determine the optimal wavelength for optically pumping Pc:Ptp in the following measurements. In terms of the photoluminescent property of Pc:Ptp, Fig.~\ref{fig:1}d indicates that the crystal can generate intense fluorescence at wavelengths of 599 and 645 nm, corresponding to the 0-0 and 0-1 transitions\cite{wang2009doped}, respectively, as well as a weak emission band central at 701 nm under illumination of a green light-emitting diode (LED). The small peak near 550 nm is attributed to the emission of LED which was not completed filtered during the fluorescence measurements. The full width at half maximum (FWHM) of the strongest fluorescence peak at 645 nm is measured to be 27.6$\pm$0.3 nm. 

\subsection{Mechanism of the simultaneous lasing and masing actions in Pc:Ptp}

Combining the optical properties obtained above with the reported properties of Pc:Ptp masers\cite{oxborrow2012room,salvadori2017nanosecond}, we propose the mechanism of realizing simultaneous lasing and masing actions by exploiting both the photoexcited singlet and triplet states of Pc:Ptp crystals. As schematically demonstrated in Fig.~\ref{fig:1}e, the pentacene molecules in the ground singlet state can be efficiently promoted to the excited singlet state with an optical pumping at 590 nm, by which the population inversion is achieved in the singlet manifold for the stimulated emission of radiation in the visible region, e.g. 599 or 645 nm where the strong fluorescence was observed. In the meantime, due to the spin-orbit coupling, pentacene molecules can also transfer to the excited triplet state (T$_{2}$) via the intersystem crossing with a yield of 62.5\% at room temperature\cite{takeda2002zero} and rapidly decay to the lowest triplet state (T$_{1}$) via the internal conversion. Since the triplet state is metastable, the pentacene molecules will eventually decay back to the ground singlet state by either the radiative or non-radiative T$_{1}\rightarrow$S$_{0}$ transition\cite{siebrand1970mechanisms,henry1971spin}. In Earth’s field, T$_{1}$ is non-degenerate and constituted by three sublevels T$_\textrm{x}$, T$_\textrm{y}$ and T$_\textrm{z}$ due to the dipolar interactions of pentacene’s triplet electron spins. The resonance frequencies of the triplet sublevels governed by the zero-field-splitting (ZFS) parameters have been determined by electron paramagnetic resonance (EPR) measurements\cite{yang2000zero,wu2019unraveling} to be 1.45 GHz, 1.344 GHz and 106.5 MHz for T$_\textrm{x}\leftrightarrow$T$_\textrm{z}$, T$_\textrm{y}\leftrightarrow$T$_\textrm{z}$ and T$_\textrm{x}\leftrightarrow$T$_\textrm{y}$ transitions, respectively. An alluring property of the pentacene’s lowest triplet state is that, upon its generation, the populations of the triplet electrons follow a non-Boltzmann distribution in the three sublevels with a ratio of $P_\textrm{x}:P_\textrm{y}:P_\textrm{z}$= 0.76:0.16:0.08\cite{sloop1981electron} at room temperature. The strong population inversion and relatively slow spin-lattice relaxation\cite{wu2019unraveling} between the T$_\textrm{x}$ and T$_\textrm{z}$ sublevels can be exploited for realizing the stimulated emission of radiation in the microwave region, i.e. masing. Therefore, the optical pumping of Pc:Ptp can simultaneously introduce population inversions in both singlet and triplet manifolds fulfilling the prerequisites of the lasing and masing actions at ambient conditions.

\subsection{Spectral analysis of the ASE of Pc:Ptp}

\begin{figure}[htbp!]
    \includegraphics[scale =0.75]{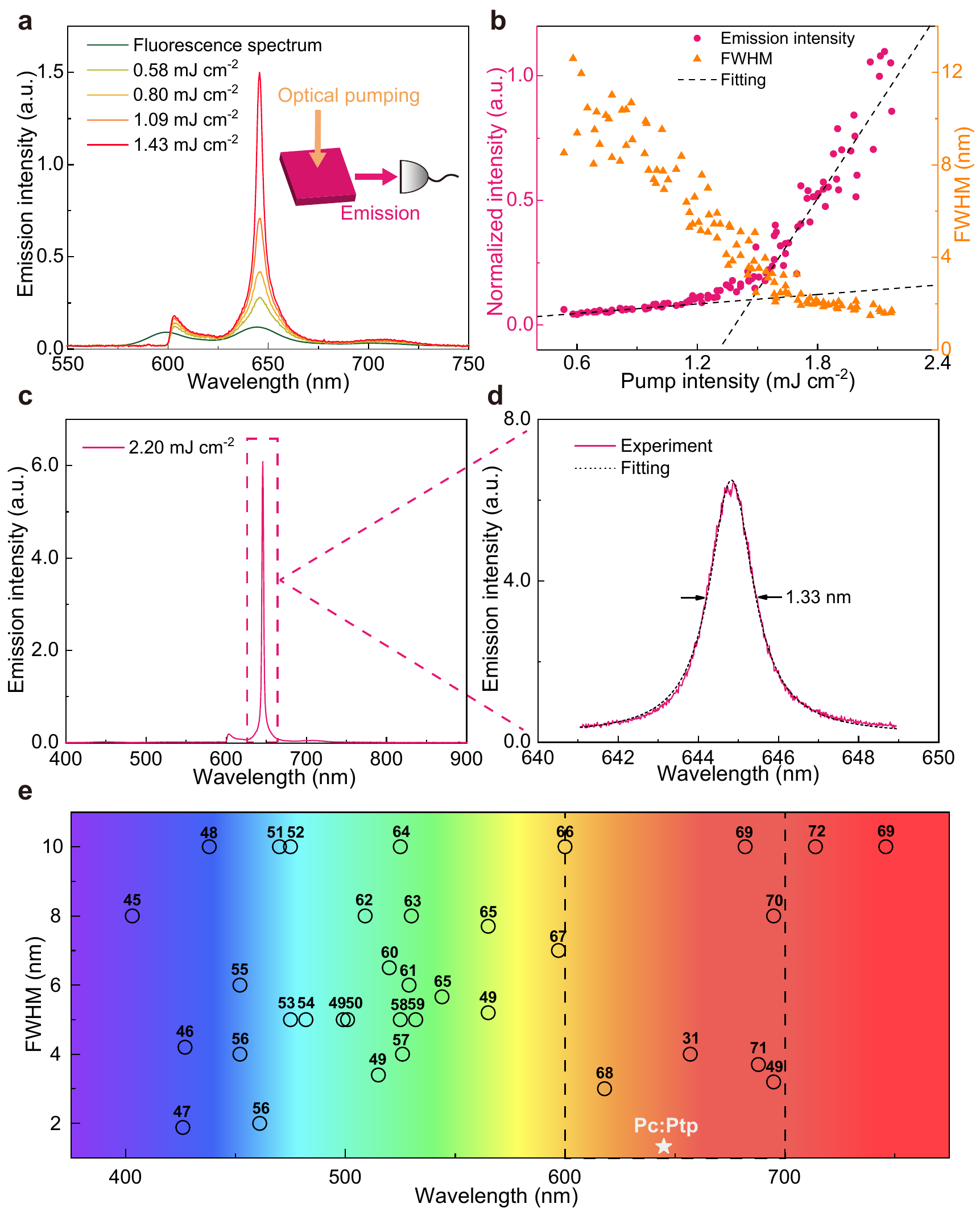}
    \caption{\textbf{Spectral analysis of Pc:Ptp's ASE process.} \textbf{a} Effect of pump intensity in the emission spectrum of Pc:Ptp. Inset: schematic of the experimental setup. \textbf{b} Dependence of the intensity (pink dots) and FWHM (orange triangles) of the emission peak at 645 nm on pump intensity. The kink behavior of the dependence reveals the ASE process whose threshold is determined by the point of intersection between the two linear fitting lines (black dotted). \textbf{c} Emission spectrum of Pc:Ptp measured above the ASE threshold. \textbf{d} Zoomed-in view of the 645-nm ASE peak (highlighted by a pink dotted box in \textbf{c}) whose FWHM is determined by a Lorentzian fitting (black dotted curve). \textbf{e} Comparison of the wavelength and FWHM of Pc:Ptp's narrowest ASE peak with those of the reported organic single crystals\cite{qu2012two,losio2007amplified,hibino2002emission,baronas2018low,ichikawa2003improved,xu2012low,xie2007amplified,park2017crystallization,wu2003structual,xu2012supramolecular,xie2007stimulated,kabe2009effect,nakanotani2009highly,mizuno2012single,xia2010efficient,chen2009two,gu2012polymorphism,chen2015low,wang2014organic,xu2010solid,sakurai2013emission,zhao2019high,fang2010two,varghese2012stimulated,ichikawa2005photopumped,cheng2015organic,fichou1997first,ichikawa2005laser,wang2014shape} reviewed in ref.\cite{jiang2020organic}. Details of the reported organic single crystals are summarized in Supplementary Information. The regime of ASE wavelength rarely reached by organic single crystals is highlighted with a black dotted box.}
    \label{fig:2}
\end{figure}

As the masing action has been successfully demonstrated with the Pc:Ptp crystal\cite{wu2021bench,wu2020room}, we verify the mechanism of multiband coherent radiation proposed above by investigating the feasibility of the stimulated emission in the pentacene’s singlet states under the experimental conditions similar to that for the maser studies. We measured the emission spectra of Pc:Ptp under the optical pumping of a nanosecond pulsed laser which has proven to be sufficiently powerful for achieving the threshold of Pc:Ptp masers\cite{salvadori2017nanosecond}. The pump laser was focused onto the cleavage plane of the crystal with a beam diameter of about 5 mm (see Supplementary Fig. 2 for the detailed experimental setup). At different pump intensities, the emission spectra shown in Fig.~\ref{fig:2}a were recorded by collecting the emitted light from the edge of the sample (see inset in Fig.~\ref{fig:2}a). It can be seen that at the peak wavelength (i.e. 645 nm) of the Pc:Ptp’s emission spectra, the intensity gradually increases while the spectral linewidth equal to the FWHM gets narrower with the increment of pump energies implying the ASE process occurs. The incomplete emission peaks at the wavelength of 600 nm is due to the long pass filter with a cut-on wavelength of 600 nm used to filter out the pump light at 590 nm. In contrast, the filter used in the fluorescence measurements has a cut-on wavelength of 550 nm leading to a complete peak at 600 nm, as shown in Fig.~\ref{fig:2}a.

To characterize the ASE process of Pc:Ptp, the intensity as well as the FWHM of the strongest emission peak at 645 nm was plotted as a function of the pump intensity as shown in Fig.~\ref{fig:2}b. There are two distinct areas in Fig.~\ref{fig:2}b, which were respectively fitted by linear equations, resulting in a kink behavior of laser-like thresholds. The difference between ASE and lasing processes is that in general, there is an optical cavity in the composition of a laser, while ASE is the stimulated emission that occurs without a cavity\cite{allen1970superradiance}. We define the kink intensity as the ASE threshold of Pc:Ptp, which is 1.47 mJ cm$^{-2}$. From the slopes of the two fitted lines, it is evident that below the threshold, with the increment of the pump intensity, the emission intensity increases slightly while the FWHM narrows rapidly, whereas the situation changes oppositely above the threshold. This is because once the pump intensity exceeds the threshold, the stimulated radiation near the wavelength of 645 nm is substantially enhanced, resulting in a rapid increase in the emission intensity in the vicinity of 645 nm manifesting a narrowing of the entire emission spectrum while the reduction of FWHM is limited by the optical inhomogeneous broadening of pentacene molecules. In Fig.~\ref{fig:2}c and \ref{fig:2}d, the narrowest emission spectrum arising from Pc:Ptp’s ASE process was obtained with a pump intensity of 2.2 mJ cm$^{-2}$. Compared with the normal fluorescence spectrum of the Pc:Ptp crystal in Fig.~\ref{fig:2}a, the ASE spectra in Fig.~\ref{fig:2}c and \ref{fig:2}d show a substantial narrowing of the linewidth from 25 to 1.33 nm, which clearly reveals the feasibility of Pc:Ptp for achieving the stimulated emission of radiation in the visible region. Most importantly, the ASE threshold determined here, i.e. 1.47 mJ cm$^{-2}$, is much lower than the maser threshold\cite{salvadori2017nanosecond} of Pc:Ptp, 26.3 mJ cm$^{-2}$ measured with the similar optical pump source, which indicates the simultaneous lasing and masing actions can be realized once the maser threshold is fulfilled.

In addition, we have also compared the ASE performance of Pc:Ptp with various organic single crystals in terms of the ASE wavelength and linewidth\cite{jiang2020organic}. In Fig.~\ref{fig:2}e, we summarized the reported organic single crystals with evident ASE behaviors of which the measured emission linewidths are below 10 nm. The types of the referred crystals can be found in the Supplementary Information. As demonstrated in Fig.~\ref{fig:2}e, the ASE wavelengths of these organic single crystals are distributed in various visible bands, especially in the wavelength range of 400-600 nm, but leave the regime between 600 and 700 nm (i.e. red-color regime) rarely reached. Thus, the ASE wavelength of Pc: Ptp at 645 nm is a good supplement to fill in this almost blank regime. Moreover, among the listed crystals, Pc:Ptp has the narrowest ASE linewidth of ~1.33 nm, as per our knowledge, which is even comparable to some organic crystal lasers\cite{li2021tunable,huang2019wavelength,zhou2020experimentally,zhang2021thermally,cai2022octagonal}. The outstanding monochromaticity reveals the potential of Pc:Ptp to be a novel organic solid-state laser gain media. 

\subsection{Kinetic analysis of the ASE of Pc:Ptp}

\begin{figure}[htbp!]
    \includegraphics[scale = 0.9]{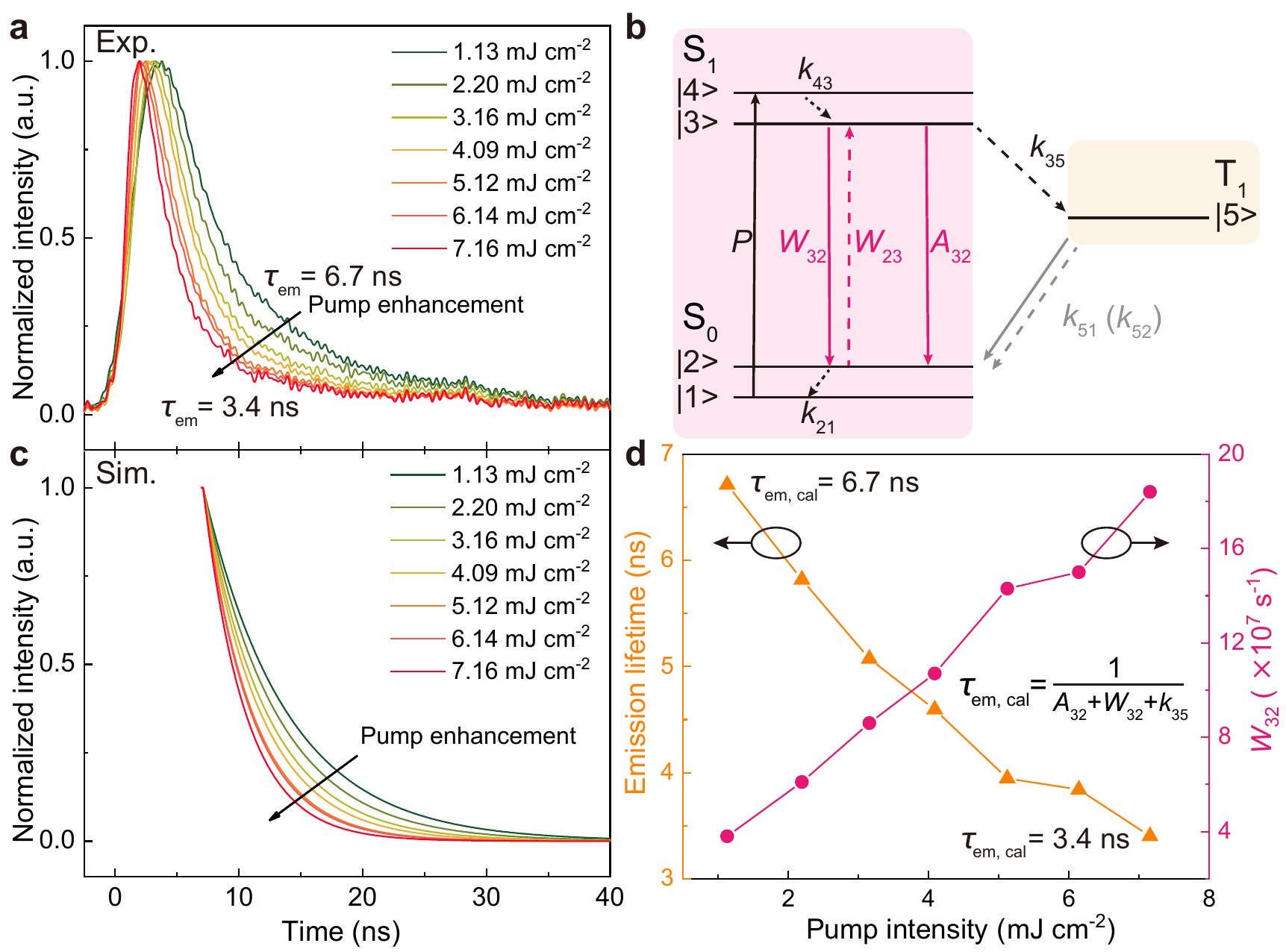}
    \caption{\textbf{Kinetic analysis of Pc:Ptp's ASE process.} \textbf{a} Pump-intensity-dependent emission decays of Pc:Ptp measured at 645 nm. The emission lifetimes are obtained with an exponential fitting. \textbf{b} Five-level kinetic model accounting for the pump-intensity-dependent emission decays of Pc:Ptp. \textbf{c} Simulation results of the pump-intensity-dependent emission decay of Pc:Ptp on the basis of the five-level model in \textbf{b}. \textbf{d} Simulated rates of stimulated emission $W_{32}$ (pink) as a function of pump intensity. The pump-intensity-dependent emission lifetimes (orange) are calculated according to the equation embedded.}
    \label{fig:3}
\end{figure}

Emission lifetimes are important parameters that can be employed to interpret the kinetic processes involved in the electronic states upon photoexcitation. We therefore further analyze the kinetic behaviors of the ASE process of Pc:Ptp based on the emission lifetime measurements. The associated experimental setup can be found in the Supplementary Information Fig. 3. As the fluorescence lifetime of Pc:Ptp has been estimated to be around 9 ns\cite{takeda2002zero} at room temperature, a photodetector with a time resolution of 1 ns resolution was employed in our setup for capturing the kinetic process accurately. Fig.~\ref{fig:3}a shows the emission decays obtained with different pump intensities. By exponential fittings of the decay curves, we found the emission lifetime was decreased from ~6.7 to 3.4 ns (as indicated by the black arrow in Fig.~\ref{fig:3}a) with enhanced optical pumping. The emission lifetime of 6.7 ns obtained with the relative weak pumping is close to the reported value of 9 ns\cite{takeda2002zero}. The faster decays observed with the stronger optical pumping implies a pump-intensity-dependent kinetic process that is included in the emission process. This behavior is consistent with the characteristic of an ASE process that the higher pump intensity will lead to enhanced stimulated emission induced by the increased photons generated from spontaneous emission. To fully characterize the observed pump-intensity-dependent emission decays, we constructed a five-level kinetic model comprising both singlet and triplet states of the pentacene molecules as demonstrated in Fig.~\ref{fig:3}b. The origins of the photoexcited singlet and triplet states are similar to that demonstrated in Fig.~\ref{fig:1}c. To reduce the complexity of the kinetic model, the numbers of the vibrational levels included in the ground (S$_{0}$) and first excited singlet states (S$_{1}$) were decreased to two, as illustrated by $\ket{1}$ and $\ket{2}$ of S$_{0}$, and $\ket{3}$ and $\ket{4}$ of S$_{1}$ in Fig.~\ref{fig:3}b. In addition, due to the extremely fast internal conversion between T$_{2}$ and T$_{1}$ in a time scale of femtosecond to picosecond\cite{iinuma1997dynamic}, the model was further simplified by assuming the direct intersystem crossing from S$_{1}$ to T$_{1}$, i.e. from the lowest vibrational level of S$_{1}$, $\ket{3}$ to $\ket{5}$ shown in Fig.~\ref{fig:3}b. Thus, the kinetic processes involved in the five-level model are the optical pumping ($\ket{1}\rightarrow\ket{4}$), the relaxation between the vibrational levels in the singlet manifold ($\ket{4}\rightarrow\ket{3}$ and $\ket{2}\rightarrow\ket{1}$), the spontaneous emission ($\ket{3}\rightarrow\ket{2}$), the simulated emission ($\ket{3}\rightarrow\ket{2}$) and absorption ($\ket{2}\rightarrow\ket{3}$) and the intersystem crossing ($\ket{3}\rightarrow\ket{5}$ and $\ket{5}\rightarrow\ket{1}(\ket{2})$).

Based on the kinetic processes, we derived a set of coupled rate equations to simulate the observed emission decays as a function of the pump intensity (see Supplementary Information). We found the simulated decay curves shown in Fig.~\ref{fig:3}c can well reproduce the measured emission decays as well as the dependence of the emission lifetimes with the pump intensity by a set of stimulated transition rates, $W_{32}$ and $W_{23}$ (see Fig.~\ref{fig:3}b and \ref{fig:3}d). The stimulated emission rate, $W_{32}$, obtained from the simulation shows an almost linear increase from $4\times10^{7}$ to $1.8\times10^{8}$ s$^{-1}$ (exceeding the spontaneous emission rate\cite{takeda2002zero}, $A_{32}= 4.2\times10^{7}$ s$^{-1}$) with the enhanced pump intensity that reveals the transition of the dominant kinetic process in the emission decay from the spontaneous emission to the stimulated emission, i.e. ASE occurs. The emission lifetimes $\tau_\textrm{em,cal}$ in Fig.~\ref{fig:3}d were calculated with $\tau_\textrm{em,cal}=\frac{1}{A_{32}+W_{32}+k_{35}}$ where $k_{35} = 6.9\times10^{7}$ s$^{-1}$ is the rate of the intersystem crossing\cite{takeda2002zero}. 

\subsection{Optimization of the ASE efficiency}

It is known that the efficiency of the transition of molecules in the ground singlet state to the excited singlet state can be maximized by aligning the polarization of pump light with the molecules’ transition dipole moments\cite{guttler1996single}. For the pentacene molecules doped in \textit{p}-terphenyl, the pentacene’s short axis (i.e. the y axis in Fig.~\ref{fig:1}b), almost parallel to the \textit{ab} cleavage plane of the crystal\cite{cui2020growth}, coincides with the transition dipole moment of the lowest spin allowed transition of pentacene\cite{pariser1956theory}. Therefore, we further attempted to optimize the ASE efficiency by enhancing the singlet transition probability of the pentacene’s molecules which would benefit the realization of a low-threshold Pc:Ptp laser in the future.

\begin{figure}[htbp!]
    \includegraphics[scale = 0.9]{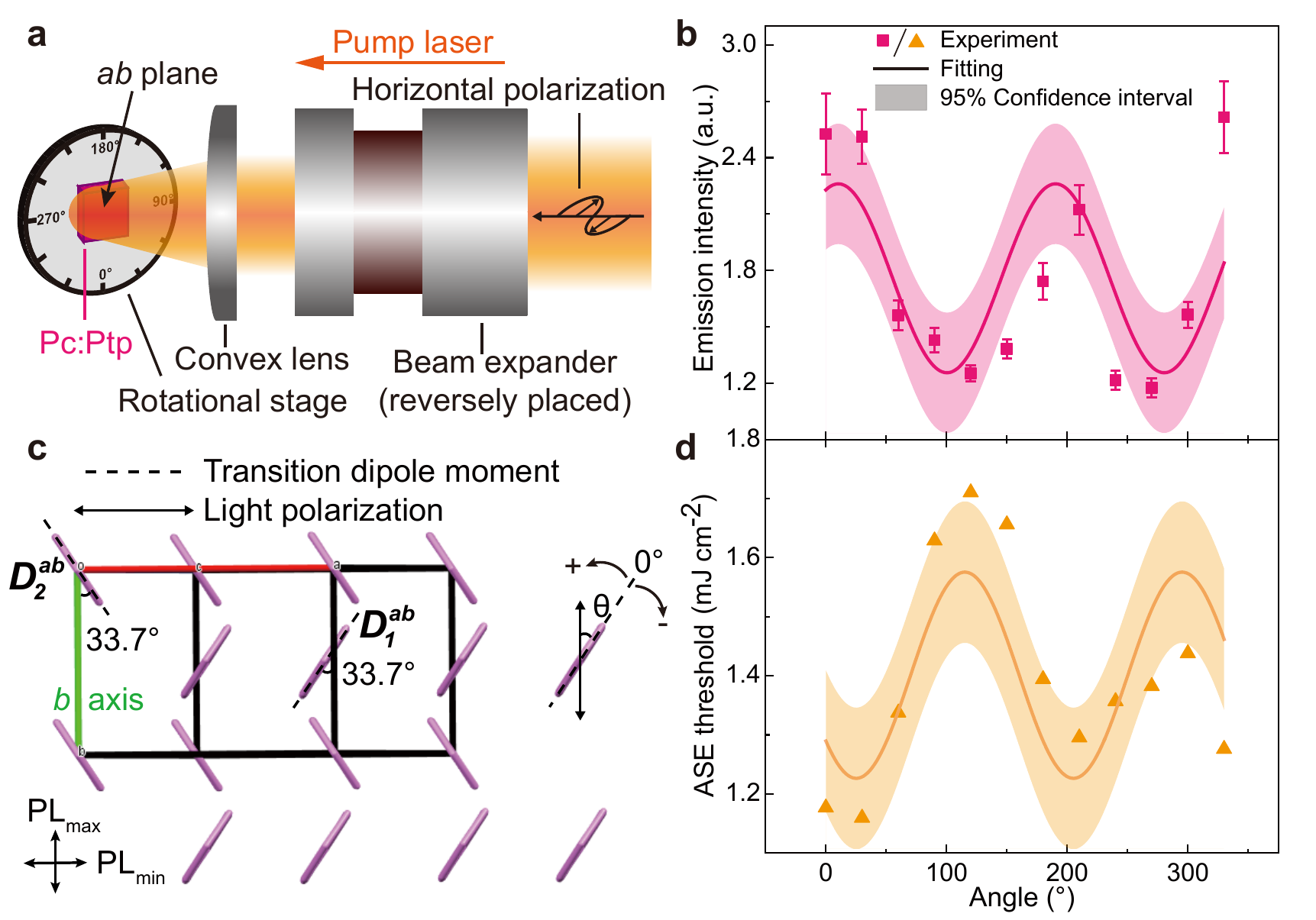}
    \caption{\textbf{Dependence of ASE performance on the alignment of crystal orientation with pump-light polarization.} \textbf{a} Schematic of the experimental setup. \textbf{b} The emission intensity (upper subplot) and ASE threshold (lower subplot) of Pc:Ptp's ASE peak at 645 nm as a function of the rotated angle of the crystal. The measured data sets (squares and triangles) are fitted by sinusoidal functions (solid curves) with a period of $180\degree$ and 95\% confidence interval bands (shadowed areas). Error bars denote the standard errors of the data. \textbf{c} Schematic diagram illustrating the orientations of pentacene's short molecular axes (purple sticks), i.e., the transition dipole moments (dashed lines) projected into the Pc:Ptp crystal's \textit{ab} plane. The angle between pentacene's short axes and light polarization (solid lines with arrows) is defined by $\theta$. The maximum and minimum emission intensities obtained respectively with $\theta = 33.7\degree$ and $-56.3\degree$ are indicated in the left bottom corner.}
    \label{fig:4}
\end{figure}

Fig.~\ref{fig:4}a schematically illustrates the experimental setup (see Methods and Supplementary Information for more details) where a horizontally polarized laser beam was focused by a combination of a reversely placed beam expander and a convex lens and propagated perpendicular with respect to the cleavage plane (i.e. \textit{ab} plane) of the Pc:Ptp crystal fixed on a rotational sample disk. Under the excitation of a fixed pump intensity exceeding the ASE threshold, the strong emission intensity resulting from the ASE process shown in Fig.~\ref{fig:4}b was measured when the crystal was rotated within the plane where the cleavage facet locates. It can be seen that the emission intensity shows an angular dependence, and the periodic behavior can be fitted by a sinusoidal function with a period of 180$\degree$, i.e. the maximum and minimum emission intensities occur with a 90$\degree$ interval. The orthogonal correlation can be explained by the convolution effect of the alignments between the light polarization and the transition dipole moments of the pentacene molecules doped in two inequivalent sites. As shown in Fig.~\ref{fig:4}c, the projections of the pentacene’s transition dipole moments into the \textit{ab} plane, $\bm{D_{i}^{ab}}$ (i = 1 and 2) are parallel to the short molecular axes of the two groups of pentacene molecules, and thus, symmetrical about the crystal \textit{b}-axis according to the room-temperature crystal structure of \textit{p}-terphenyl\cite{rietveld1970x}. The angles of the two transition dipole moments with respect to the \textit{b}-axis are both 33.7$\degree$. By assuming the transition dipole moments of the two groups of pentacene molecules only vary in terms of the orientation, the obtained emission intensity is proportional to $\bm{|D_{1}^{ab} \cdot E|}^{2}+\bm{|D_{2}^{ab} \cdot E|}^{2}$ where $\bm{E}$ is the electric field vector of the laser light in the \textit{ab} plane\cite{guttler1993single}. By denoting the angle between the light polarization and one of the transition dipole moments to be $\theta$ ($-90\degree < \theta \leq 90\degree$), the emission intensity is found to be proportional to $1+\cos[{(\frac{2\theta}{180}\pi)-\frac{67.4}{180}\pi}]\cos{(\frac{67.4}{180}\pi)}$ which implies a modulation of the emission intensity with a period of 180$\degree$ and matches with our measurements. It can also be found that the maximum and minimum emission intensities should be obtained when $\theta=33.7\degree$ and $-56.3\degree$ which correspond to the scenarios where the light polarization is parallel and perpendicular to the \textit{b} axis, respectively.

Moreover, the ASE threshold was measured as a function of the rotation angle which reveals a similar periodic trend and orthogonal relationship but with a $\sim90\degree$ offset of the extreme points with respect to those in Fig.~\ref{fig:4}d. This offset is due to that the highest singlet transition probability indicated by the strongest emission intensity in Fig.~\ref{fig:4}d facilitates the buildup of the population inversion in the singlet manifold and thus reduces the threshold of achieving the ASE process, and vice versa.

Therefore, by adjusting the angle of the light polarization with respect to the pentacene’s transition dipole moment, the highest singlet transition probability can be achieved, offering the advantages of a two-fold enhancement of the emission intensity and a reduction of the ASE threshold by around 30\% (see Fig.~\ref{fig:4}b and \ref{fig:4}d) compared to those measured at an orthogonal position. This strategy will be beneficial for lowering not only the lasing threshold of Pc:Ptp, but also its masing threshold, because the facilitated transition to the excited singlet state can also lead to the more efficient generation of the pentacene’s triplet states via the intersystem crossing.

\section{Discussion}

In summary, our study reveals the unexplored potential of Pc:Ptp crystals as room-temperature laser gain media which has been overlooked in previous fluorescent and magnetic-resonance spectroscopic studies\cite{sloop1981electron,guttler1996single,kohler1999magnetic} on Pc:Ptp’s photoexcited spin states. Even without an optical cavity, the stimulated emission observed at 645 nm with a narrow linewidth around 1 nm shows a great promise of Pc:Ptp lasers to fill the wavelength gap of the existing organic solid-state lasers. Most importantly, since Pc:Ptp masers have been realized with the identical crystals and optical-pumping conditions\cite{salvadori2017nanosecond}, our findings prove the feasibility of achieving simultaneous lasing and masing actions with Pc:Ptp crystals at room temperature.

The next step will be to fabricate a multiband coherent device by incorporating a Pc:Ptp crystal with a hybrid cavity architecture supporting both resonances at 645 nm and 1.45 GHz. Considering the volumes of the three-dimensional (3D) dielectric microwave cavities\cite{oxborrow2012room,breeze2015enhanced} and the Pc:Ptp crystals employed in the Pc:Ptp masers, a Fabry-P\'erot optical cavity could be a compatible choice for promoting the lasing action while not perturbing the microwave electromagnetic modes in the 3D dielectric cavities. The pumping threshold of the multiband coherent radiations can be minimized by appropriate alignments of the pentacene’s short molecular axis with the polarization of the optical pumping, as well as the magnetic field of the electromagnetic mode in the microwave cavity\cite{breeze2015enhanced}. We envision that the correlation and manipulation of the optical and microwave photons simultaneously generated by the proposed multiband coherent source are worth being investigated for fundamental tests of quantum optics, the possibility of phase locking for development of self-referenced frequency combs  and optimization of the solid-state quantum sensors exploiting the nonlinear behaviors of the stimulated emission in either the microwave\cite{wu2022enhanced} or visible\cite{hahl2022magnetic} band.

\section{Methods}

\subsection{Sample preparation}

A Pc:Ptp single crystal with a doping concentration of 1000 p.p.m. was grown with the Bridgman method as reported in ref.\cite{oxborrow2012room}. The as-grown Pc:Ptp crystal was cut to obtain a cleavage facet which was successively polished by abrasive papers, 0.1-$\mu$m cerium oxide powder and 0.05-$\mu$m aluminum oxide powder. The surface parallel to the finished facet was polished by repeating the above procedures.

\subsection{Optical characterizations}

The UV/vis absorption spectrum of Pc:Ptp was collected using a UV-visible-near infrared spectrophotometer (Lambda 1050+, PerkinElmer). The fluorescence spectrum of Pc:Ptp was collected using a home-built setup whose block diagram is shown in Supplementary Information Fig.1. A green LED source was used to illuminate the sample, and the fluorescence spectrum was collected by an optical spectrum analyzer (Maya 2000 Pro, Ocean Optics, resolution 1 nm). The optical microscopic images were taken by a Complementary Metal-Oxide-Semiconductor Transistor (CMOS) camera (AP-MV-UH1080, Apico).

\subsection{ASE measurements}

The ASE properties of Pc:Ptp were determined via a home-built setup illustrated in Supplementary Information Fig.2. An optical parametric oscillator (OPO) (BBOPO-Vis, Deyang Tech. Inc., pulse duration 7 ns) pumped by an Nd:YAG Q-switched laser (Nimma-900, Beamtech, repetition rate 10 Hz) with horizontal polarized output at 590 nm was used for the ASE measurements. The OPO output beam was focused on the sample surface by a reversely placed beam expander (2x) and a convex lens with a focal length of 20 cm. The beam diameter was 5 mm, which completely covered the sample surface. A 50/50 beam splitter was used to divert the pump light for measuring the pump energy with a energy meter (BGS6321, Beijing Institute of Optoelectronic Technology). A long-pass filter with cut-on wavelength of 600 nm was used to eliminate the pump light. The ASE signals were collected by an optical fiber connected to a high-resolution spectrometer (SpectraPro HRS-750, Pro EM 512B, Teledyne Princeton Instruments).

\subsection{Emission lifetime measurements}

The experimental setup was similar to that used for the ASE measurements except the spectrometer was replaced by a photodetector (DET10A2, Thorlabs, resolution 1 ns). The time-domain emission signals under several different pump energies were collected by an oscilloscope (WAVERUNNER 6KA, LeCroy).

\subsection{Orientation-dependent emission measurements}

The setup was the same as the ASE measurements. The Pc:Ptp crystal was fixed on the center of a rotational disk (HRSP40-L, Heng Yang Optics, $0.1\degree$ resolution) so that the incident light can propagate perpendicular to the cleavage plane. The crystal was rotated with an interval of $30\degree$ between each measurement. The emission spectra of Pc:Ptp were measured at different rotation angles under the same pump intensity of 1.53 mJ cm$^{-2}$. The ASE thresholds were measured by varying the pump intensity at different rotation angles with an interval of $30\degree$.

\begin{acknowledgement}

The authors sincerely thank Shamil Mirkhanov for stimulating discussions and Tan Wang and FORTEC Technology (HK) Co.Ltd. for providing us with the monochromator SpectraPro HRS-750. H.W. acknowledges financial support from the National Science Foundation of China (NSFC) (12204040) and the China Postdoctoral Science Foundation (YJ20210035 and 2021M700439). Jiyang Ma acknowledge financial support from China National Postdoctral Program for Innovative Talents (BX20200057). 

\end{acknowledgement}





\providecommand{\latin}[1]{#1}
\makeatletter
\providecommand{\doi}
  {\begingroup\let\do\@makeother\dospecials
  \catcode`\{=1 \catcode`\}=2 \doi@aux}
\providecommand{\doi@aux}[1]{\endgroup\texttt{#1}}
\makeatother
\providecommand*\mcitethebibliography{\thebibliography}
\csname @ifundefined\endcsname{endmcitethebibliography}
  {\let\endmcitethebibliography\endthebibliography}{}

\end{document}